\documentstyle[epsfig,aps,fleqn,preprint,tighten]{revtex}
\begin{document}
\draft
\title{Arbitrary Phase Rotation of the Marked State Can not Be Used for
Grover's Quantum Search Algorithm}
\author{Gui Lu Long, Wei Lin Zhang, Yan Song Li and Li Niu}
\address{Department of Physics, Tsinghua University, Beijing, 100084, P.
R. China}
\date{April 9, 1999}
\maketitle
\begin{abstract}
A misunderstanding that  an arbitrary phase rotation of the marked state
together with the inversion about average operation in
Grover's search algorithm can be used to construct a (less efficient) 
quantum search
algorithm is cleared. The
$\pi$ rotation of the phase of the marked state is not only the choice
for efficiency, but also vital in Grover's quantum search algorithm. The
results also show that Grover's quantum search algorithm is robust.
\end{abstract}


Grover's quantum search algorithm is one of the most important
development in quantum computation\cite{r1}. It achieves quadratic
speedup in searching a marked state in an unordered list over classical
search algorithms. As the algorithm involves only simple operations, it
is easy to implement in experiment. By now, it has been realized in NMR
quantum computers\cite{r2,r3}. Benett et al \cite{r4} have shown that no
quantum 
algorithm can solve the search problem in less than $O\sqrt{N}$ steps.
Boyer et al \cite{r5} have given analytical expressions for the
amplitude of the states in Grover's search algorithm and given tight
bounds. Zalka\cite{r6} has improved this tight bounds
and showed that Grover's algorithm is optimal. Zalka
also proposed\cite{r7} an improvement on Grover's algorithm. In another
development, Biron et al\cite{r8} generalized Grover's algorithm to an
arbitrarily distributed initial state. Pati\cite{r9} recast the
algorithm in geometric language and studied the bounds on the algorithm.

In each iteration of the Grover's search algorithm, there are two steps:
1) a selective inversion of the amplitude of the marked state, which is
a phase rotation of $\pi$ of  the marked state; 2) an inversion about
the average of the amplitudes of all basis states. This second step can
be realized by two Hadamard-Walsh transformations and an rotation of
$\pi$ of the all basis states different from $|0\rangle$. 
Grover's search
algorithm is a series of rotations in an SU(2) space span by
$|n_0\rangle$, the marked state and $|c\rangle={1\over
\sqrt{N-1}}\sum_{n\neq n_0}|n\rangle$. Each iteration rotates the state
vector of the quantum computer system an angle $\psi=2\arcsin{1\over
\sqrt{N}}$ towards the $|n_0\rangle$ basis of the SU(2) space.
Grover further showed\cite{r10} that the 
Hadamard-Walsh transformation can be replaced by almost any unitary
transformation. The inversions of the amplitudes can be instead
rotated by arbitrary phases\cite{r10}. 
It is believed that\cite{r10,r7} if one rotates the phases of the
states arbitrarily,
the resulting transformation is still a rotation 
of the state
vector of the quantum computer towards the $|n_0\rangle$ basis in the SU(2)
space. But the angle of rotation is smaller than $\psi$. From the
consideration of efficiency, the phase rotation of $\pi$ should be
adopted. This fact has been used to the advantage 
by Zalka recently\cite{r7} to improve
the
efficiency of the quantum search algorithm. 
According to the proposal,  
the inversion of the amplitude of the marked state 
in step 1 is replaced by a rotation through an angle between 0 and
$\pi$ to produce a smaller angle of SU(2) rotation towards the end of a
quantum search calculation so that the amplitude
of the marked state in the computer system state vector is exactly 1.

In this Letter, we show by explicit construction that the above concept
is actually wrong. When the rotation of the phase of the marked state is
not $\pi$, one can simply not construct a quantum search algorithm at
all. Suppose the initial state of the quantum computer is
\begin{eqnarray}
|\phi\rangle=B_0|n_0\rangle+A_0{1\over \sqrt{N-1}}\sum_{n\neq
n_0}|n\rangle.
\label{e1}
\end{eqnarray}
The modified quantum search algorithm now consists of the following two
steps:
1) $|n_0\rangle\rightarrow e^{i\theta}|n_0\rangle$; 2) an inversion
about the average operation $D$, whose matrix elements are:
\begin{eqnarray}
D_{ij}=\left\{
\begin{array}{ll}{{2\over N},} & {i\neq j}\\
                            {{2\over N}-1,}& {i=j}
\end{array}\right.
\label{e2}
\end{eqnarray}
After each iteration of the modified Grover's quantum search, the state
vector still has the form of (\ref{e1}). The recurrent formula for the
amplitudes are
\begin{eqnarray}
B_{j+1}&=&-{N-2 \over N}e^{i\theta}B_j+{2\sqrt{N-1}\over
N}A_j,\nonumber\\
A_{j+1}&=&{2\sqrt{N-2} \over N}e^{i\theta}B_j+{N-2\over N}A_j.
\label{e3}
\end{eqnarray}
Denoting $\cos\psi={N-2\over N}$, $\sin\psi={2\sqrt{N-1}\over N}$, we
can rewrite the recurrent relation in matrix form:
\begin{eqnarray}
\left(\begin{array}{c} B_{j+1} \\ A_{j+1}\end{array}\right)
=\left(\begin{array}{rr} -\cos\psi e^{i\theta} & \sin\psi \\
                       \sin\psi e^{i\theta} & \cos\psi\end{array}\right)
\left(\begin{array}{c} B_{j} \\ A_{j}\end{array}\right).
\label{e4}
\end{eqnarray}
It is not difficult to diagonalize the transformation matrix. The
eigenvalues are:
\begin{eqnarray}
\lambda_{1,2}=e^{i\gamma_{1,2}},
\end{eqnarray}
with
\begin{eqnarray}
\sin\gamma_{1,2}={-\sin\theta\cos\psi\pm
2\sqrt{1-\cos\psi^2\sin^2\theta}\sin{\theta\over 2} \over 2}.
\end{eqnarray}
It is worth pointing that the two eigen-phases satisfy
$\gamma_1+\gamma_2=\pi+\theta$. The corresponding normalized
eigenvectors are the column vectors of the matrix $U$,
\begin{eqnarray}
U=\left(\begin{array}{cc}{\sin\psi \over
\sqrt{2(1-\cos\psi\cos\gamma_2)}} & {-\cos\psi+e^{i\gamma_2} \over
\sqrt{2(1-\cos\psi\cos\gamma_2)}}\\
{\cos\psi e^{i\theta}+ e^{i\gamma_1} \over
\sqrt{2(1-\cos\psi\cos\gamma_2)}} &
{\sin\psi e^{i\theta} \over
\sqrt{2(1-\cos\psi\cos\gamma_2)}}\end{array}\right). 
\label{e6}
\end{eqnarray}
This $U$ matrix is unitary and diagonalizes the transformation matrix in
(\ref{e4}), that is  $U^{-1}TU$ is diagonal.
The amplitude of the marked state after $j+1$ iterations is
\begin{eqnarray}
B_{j+1}&=&{\sin\psi \over 2(1-\cos\psi\cos\gamma_2)}
e^{i(j+1)\gamma_1}\left[
\sin\psi B_0+(\cos\psi e^{-i\theta}+e^{-i\gamma_1})A_0\right]\nonumber\\
&&+{-\cos\psi+e ^{i\gamma_2} \over 2(1-\cos\psi\gamma_2)}
e^{i(j+1)\cos\gamma_2} \left[(-\cos\psi+e^{-i\gamma_2})B_0+\sin\psi
e^{-i\theta} A_0\right].
\label{e7}
\end{eqnarray}
When $\theta=\pi$ and $B_0=\sqrt{1\over N}$, $A_0=\sqrt{N-1\over N}$, we
recover the original Grover's quantum search algorithm, and 
$B_{j+1}=\sin((j+1+1/2)\psi)$ as given by Boyer et al\cite{r5}.

To see the effect of the rotation angle $\theta$ on the quantum search
algorithm, we plot the norm $|B_{j+1}|$ with respect to $\theta$. As
examples, we draw
in Fig. 1. $|B_4|$, and $|B_7|$ in Fig. 2. For simplicity, $N=100$,
$B_0=\sqrt{1\over N}$ and $A_0=\sqrt{N-1\over N}$. From these studies,
we see the following points: 

1) as $j$ increases, $|B_j|$ increases too
for small $j$ values for $\theta=\pi$. When  $\theta=\pi$, 
Grover's original quantum search algorithm is working. 

2) For other values of $\theta$ between 0 and $2\pi$, the dependence of
$|B_{j+1}|$ on $\theta$ is not monotonic. There are oscillations. There are
peaks and valleys in the values of $|B|$ for a given $j$. 
What is more, when $j$
changes, the positions of these peaks and valleys change too. In other
words, at a given $\theta$ value, $|B_{j+1}|$ 
does not always increase when $j$
increases. For instance, when $j=3$, there is only one peak for $\theta$
between 0 and $\pi$, whereas for $j=6$, there are 3 peaks.
This is contrary to the common expectation that for small
number of iterations, $|B_{j+1}|$ should monotonically 
increase, though not as
big as the standard Grover's quantum search algorithm.

3) For a $\theta$ different from $\pi$, even one increases the number of
iterations, the norm of the amplitude of the marked state can not reach
one. There is a limit at which the norm of the amplitude can reach. In
Fig 3. and Fig. 4, we plot the $|B_{j+1}|$ versus $j$ for $\theta={\pi
\over 4}$ and $\theta={\pi \over 3}$ respectively. The behavior is quite
interesting. For $\theta={\pi \over 4}$, there is rapid irregular
oscillations in the norm. In  particular, the maximum height is only
about 0.15. The minimum is not zero, it is about 0.07.
For $\theta={\pi\over 3}$,  the plot can be seen as  three
lines at an interval of 3 points. Again, the maximum height is
small, only about 0.18. The norm of the amplitude is in a range between
0.06 and 0.18. Even if one increases the number of iterations,
the norm can not be increased any further. In this case, we have ploted
$j$ up to 100, which is equal to the number of items in the unsorted
system. 

4) In the vicinity of $\pi$, the algorithm still works, though the
height of the norm can not reach 1. But it can still reach a considerably
large
value.  This shows that Grover's quantum search algorithm is robust
with respect to $\theta$ at $\pi$. This is important as an imperfect gate
operation may lead to a phase rotation not exactly equal to $\pi$.
Grover's quantum search algorithm has  a good tolerance on the phase
rotating angle near $\pi$. A small deviation from $\pi$ 
will not destroy the algorithm. 

To summarize, we see that $\theta=\pi$ is not only a requirement for
efficiency, but also a necessary condition for the algorithm. At this
angle, the algorithm is also robust. To achieve a smaller increase in
the marked state amplitude(or a smaller rotation towards the marked
state basis in the SU(2) space), one has to resort to more complicated
modifications to Grover's quantum search algorithm.

Encouragement from Prof. Prof. Haoming Chen is
gratefully acknowledged. We thank  Prof. Grover for helpful email 
discussions regarding Grover's quantum search algorithms and bringing
our attention new references on the algorithm.

\begin{figure}
\begin{center}
\epsfig{figure=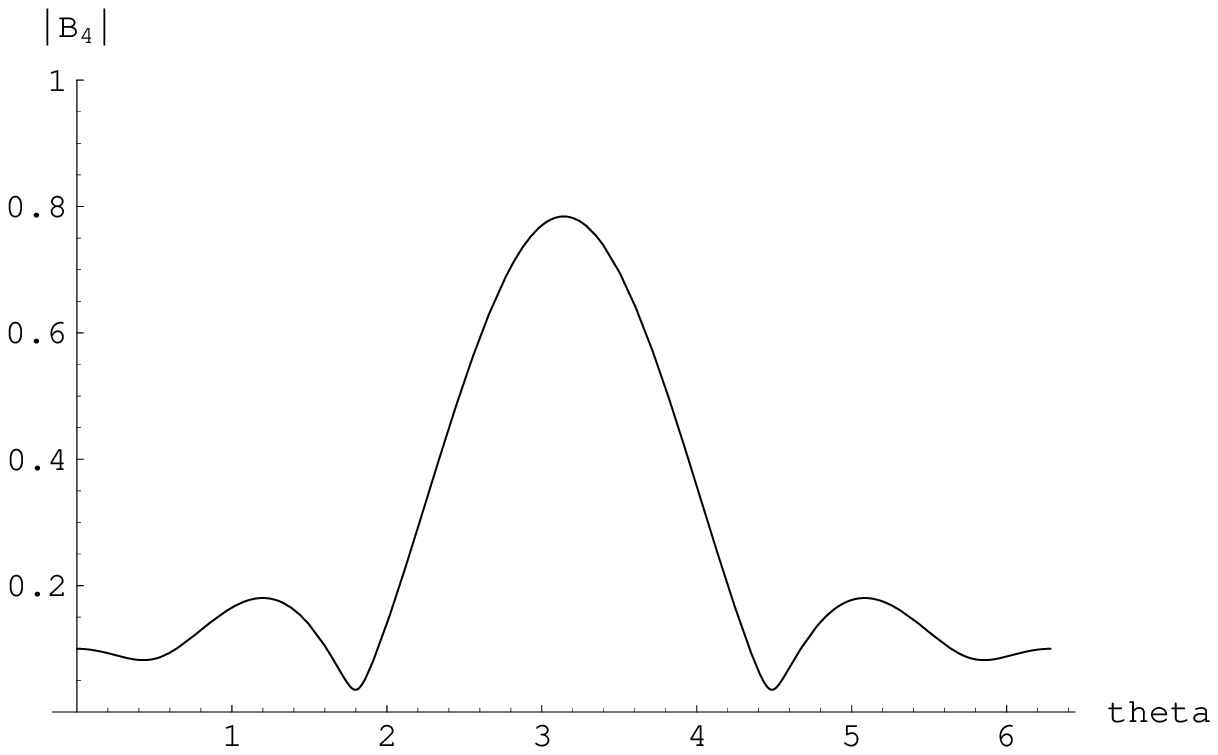,width=10cm}
\caption{$|B_{4}|$ versus $\theta$.}
\end{center}
\end{figure}
\begin{figure}
\begin{center}
\epsfig{figure=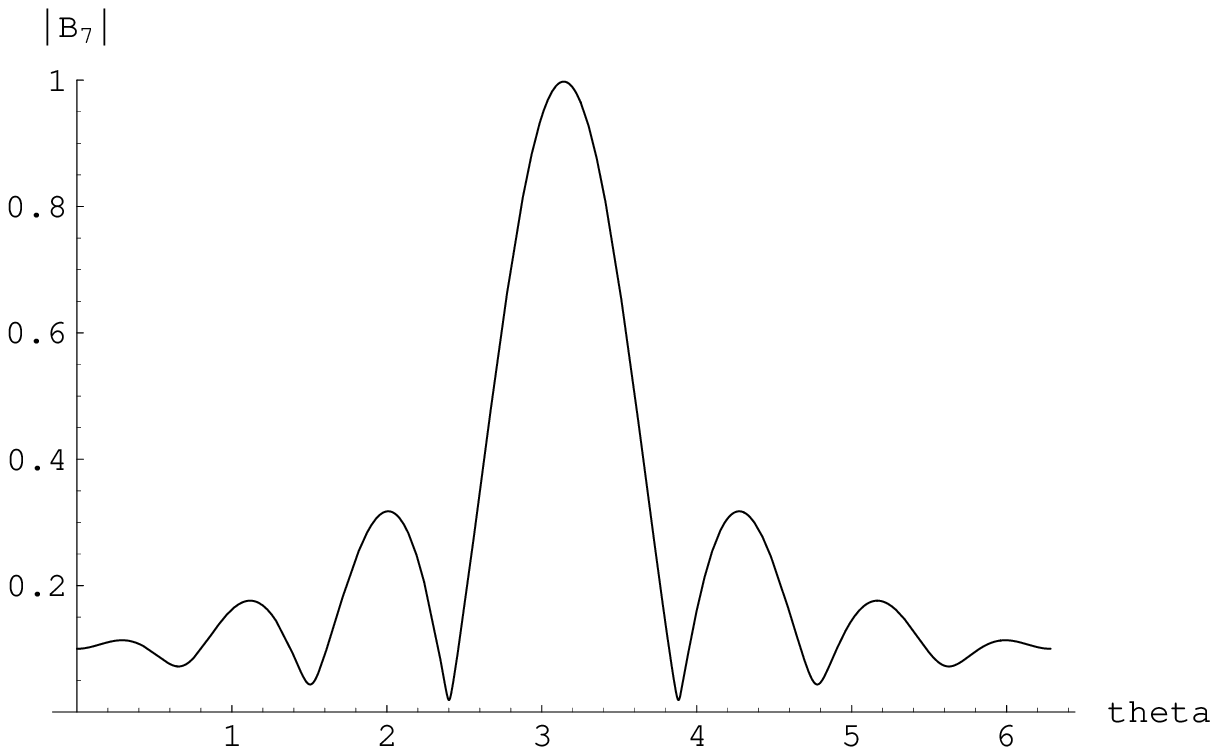,width=10cm}
\caption{$|B_{7}|$ versus $\theta$.}
\end{center}
\end{figure}

\begin{figure}
\begin{center}
\epsfig{figure=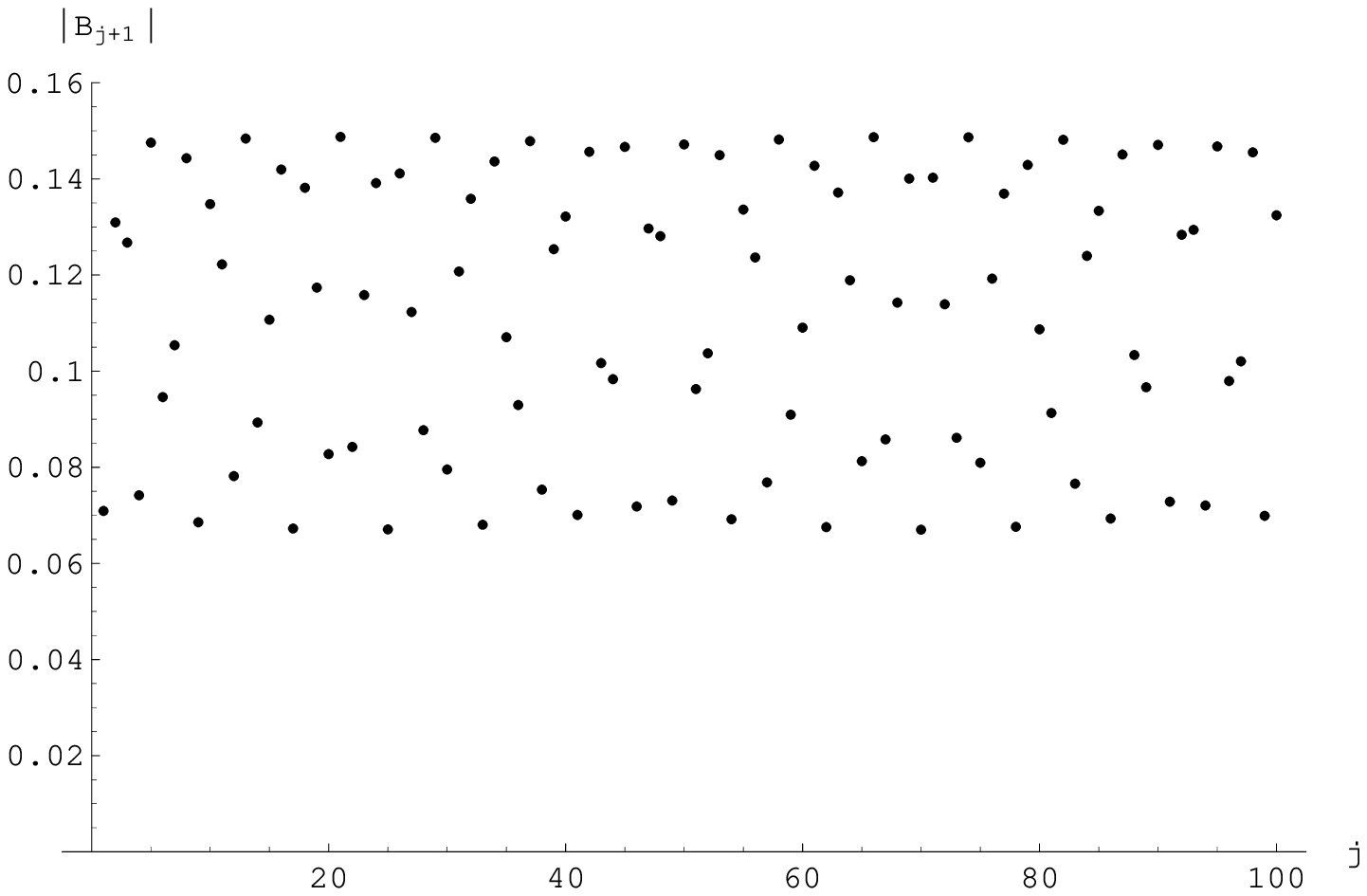,width=10cm}
\caption{$|B_{j+1}|$ versus $j$ for $\theta={\pi\over 4}$.}
\end{center}
\end{figure}

\begin{figure}
\begin{center}
\epsfig{figure=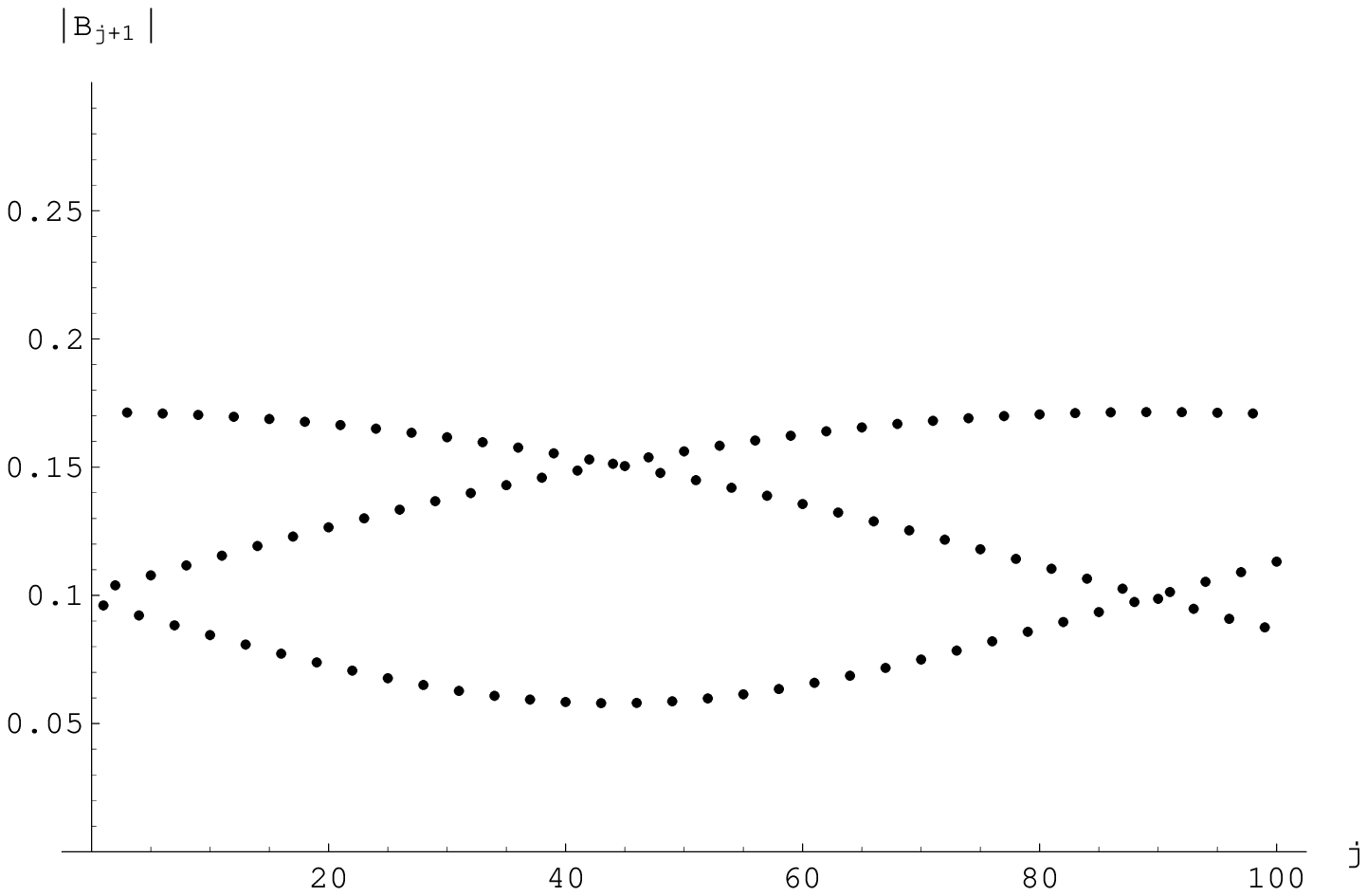,width=10cm}
\caption{$|B_{j+1}|$ versus $j$ for $\theta={\pi\over 3}$.}
\end{center}
\end{figure}

\begin{figure}
\begin{center}
\epsfig{figure=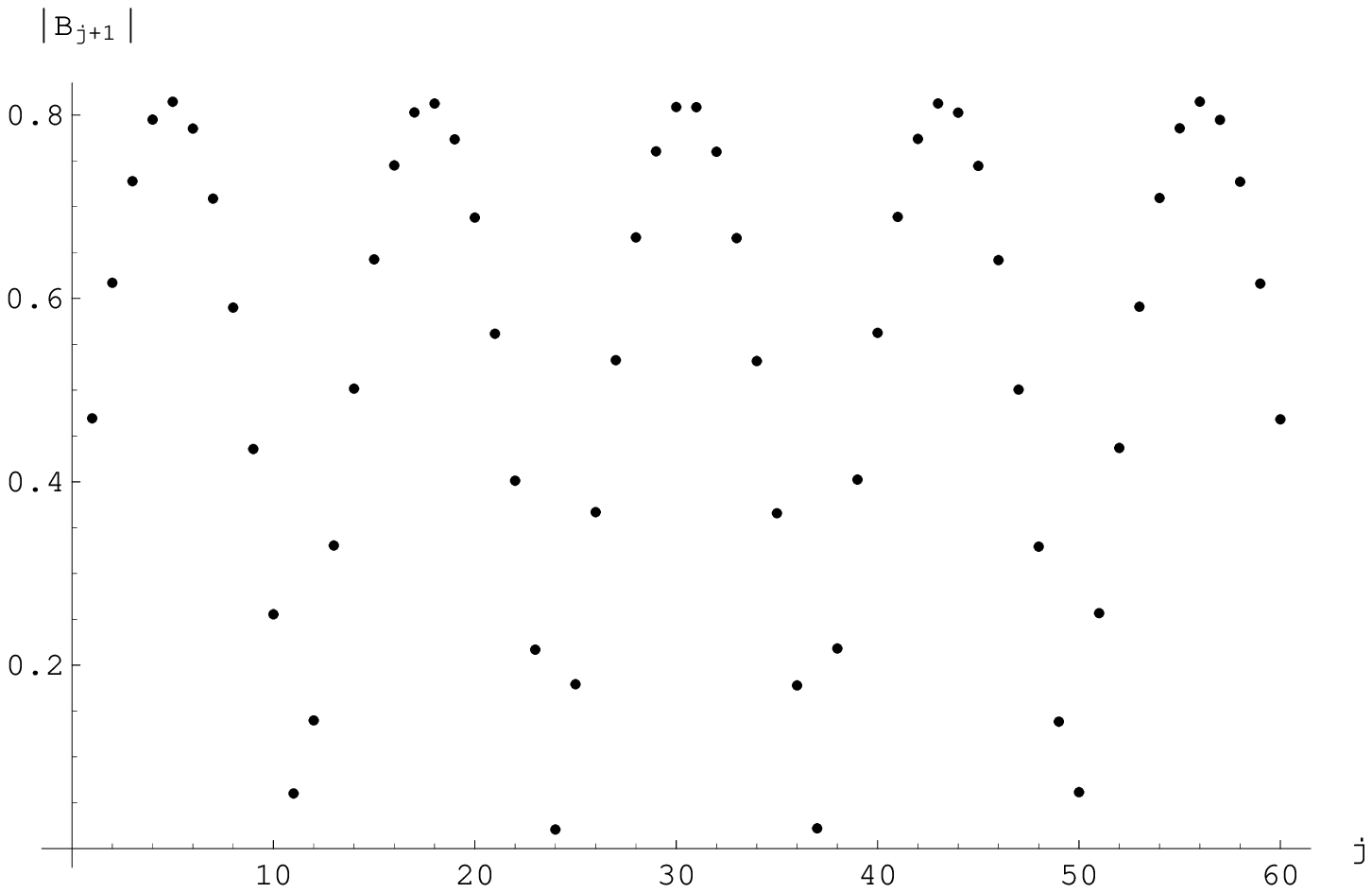,width=10cm}
\caption{$|B_{j+1}|$ versus $j$ for $\theta={\pi\over 1.1}$.}
\end{center}
\end{figure}

\end{document}